\documentclass[
 reprint,
 amsmath,amssymb,
 aps,
 prl,
 showpacs,
 superscriptaddress
]{revtex4-1}

\usepackage{graphicx}
\usepackage{dcolumn}
\usepackage{bm}
\usepackage{dsfont}
\usepackage[separate-uncertainty=false]{siunitx}

\usepackage{braket}
\usepackage{booktabs}
\usepackage{tabu}

\begin{document}

\title{A double-well atom trap for fluorescence detection at the Heisenberg limit}

\author{Ion Stroescu}
\email{selectivecounting@matterwave.de}
\affiliation{Kirchhoff-Institute for Physics, University of Heidelberg, INF 227, 69120 Heidelberg, Germany}

\author{David B. Hume}
\affiliation{Kirchhoff-Institute for Physics, University of Heidelberg, INF 227, 69120 Heidelberg, Germany}
\affiliation{National Institute of Standards and Technology, 325 Broadway, Boulder, CO 80305, USA}

\author{Markus K. Oberthaler}
\affiliation{Kirchhoff-Institute for Physics, University of Heidelberg, INF 227, 69120 Heidelberg, Germany}


\date{\today}

\begin{abstract}
We experimentally demonstrate an atom number detector capable of simultaneous detection of two mesoscopic ensembles with single-atom resolution. Such a sensitivity is a prerequisite for quantum metrology at a precision approaching the Heisenberg limit. Our system is based on fluorescence detection of atoms in a novel hybrid trap in which a dipole barrier divides a magneto-optical trap into two separated wells. We introduce a noise model describing the various sources contributing to the measurement error and report a limit of up to 500 atoms for single-atom resolution in the atom number difference.
\end{abstract}

\maketitle

\section*{Introduction}

Single-particle resolution in atom number detection, i.\,e.~atom counting, represents the ultimate limit in detector efficiency. This level of resolution is needed for observing a variety of quantum effects, and, perhaps more so, for using those quantum effects in areas such as quantum-enhanced metrology~\cite{Giovannetti:2004jg} and quantum simulation. A paradigm example is metrology at the Heisenberg limit, where the phase precision in a measurement with $N$ atoms is given by $\Delta \phi \sim 1/N$. A general model of such a measurement, describing the most common schemes such as Ramsey spectroscopy and spatial atom interferometry, is a coupled two-mode system. Here, $N$ atoms enter a beam splitter, populating the two modes, evolve through the two arms of the interferometer, then recombine at the second beamsplitter, producing two output modes where the atom number is detected independently. The ideal detector, enabling phase resolution at the Heisenberg limit, determines the exact atom number in each mode simultaneously. The atom number difference of the two modes is relevant for standard interferometry and the sum for SU(1,1) interferometers~\cite{Yurke:1986uf}. For mesoscopic ensembles of hundreds of atoms, such a capability has not been realized experimentally.

\begin{figure}[b]
\includegraphics[width=0.5\textwidth]{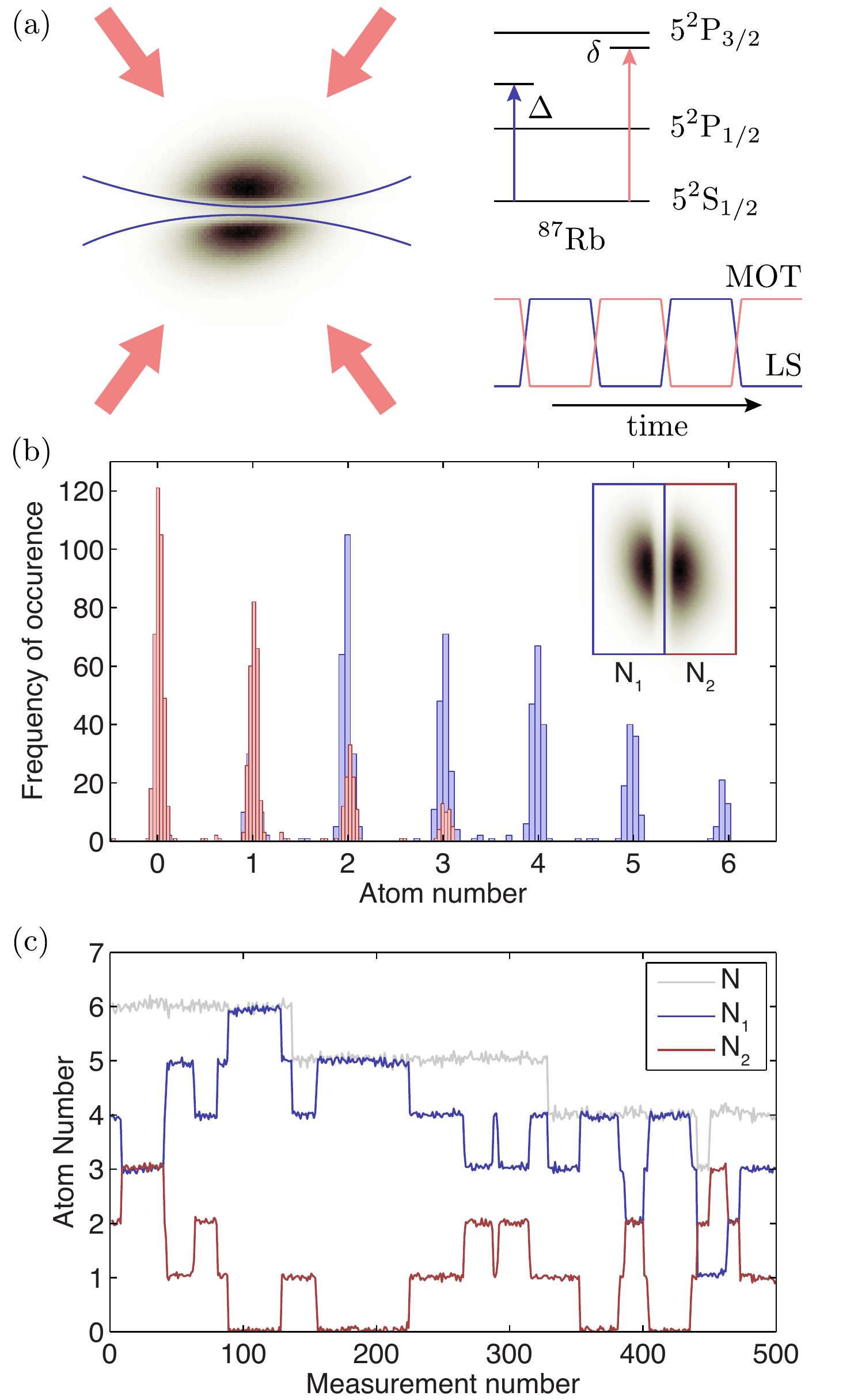}
\caption{(Color online) Fluorescence imaging in a split magneto-optical trap. (a) A blue-detuned light-sheet divides the MOT into two separate regions. The level scheme shows the laser cooling detuning $\delta$ and the light-sheet detuning $\Delta$. The two light sources are alternately pulsed. (b) The histograms of the individual fluorescence signals from each site of the split MOT reveal a very clear separation of the calibrated atom numbers $N_1$ and $N_2$. (c) At about half the maximum barrier height the individual time traces show anti-correlated hopping events, while the total atom number changes considerably less.}
\label{fig:Figure1}
\end{figure}

In recent years, spin-squeezing and entanglement in ensembles of neutral atoms has been demonstrated~\cite{Esteve:2008ij,Appel:2009jr,Leroux:2010jg,Haas:2014kb}, as well as the resulting improvement of phase sensitivity in atom interferometry~\cite{Gross:2010jn,Riedel:2010ge,Bohnet:2014}. The techniques to produce and analyze these states continue to be developed toward greater entanglement, higher atom numbers, and ultimately toward real applications. In the majority of these demonstration experiments, detection noise is already a limiting factor, and the true quantum resource provided must be inferred based on calibrating this technical noise out of the measurement. When using the atoms as quantum sensors in a metrological setting, noise subtraction is not possible, because technical noise is indistinguishable from quantum projection noise. As the techniques for producing entangled states of atoms improve and applications are developed for quantum enhanced sensors, more demand will be placed on detector precision. Here, we demonstrate a method of simultaneous fluorescence detection of two spatially-separated atomic ensembles with single-atom resolution, suitable for reaching Heisenberg-limited interferometry with many hundreds of atoms.

Single-atom resolution has been demonstrated previously via fluorescence detection of small atom numbers in experiments with magneto-optical traps~\cite{Hu:1994vq,Willems:1997uo,Gomer:2001ub,Haubrich:1996ue} and optical dipole traps~\cite{Karski:2009dv,Gibbons:2011je,Fuhrmanek:2011ea} as well as freely propagating atoms~\cite{Bucker:2009kl,Fuhrmanek:2010fs}. Single-atom sensitivity has also been achieved by monitoring light either reflected or transmitted through an optical cavity~\cite{Puppe:2007fx,Poldy:2008dg,Gehr:2010eh}. Detecting larger numbers of atoms at the single-atom level is more difficult for several reasons. First, the higher signal levels, are accompanied by higher noise. For example in fluorescence detection of $N$ atoms, photon shot noise, which must remain less than the signal from a single atom, scales with $\sqrt{N}$. Second, the probability of losing a single atom just before or during detection scales with $N$. In fact, interactions between atoms often lead to worse scaling of the loss with atom number. Nevertheless, impressive atom number resolution has been achieved in experiments with mesoscopic atom numbers. A common detection technique for Bose-condensed atoms is absorption imaging~\cite{Muessel:2013hs}, which has been optimized~\cite{Ockeloen:2010bv} to a resolution of about four atoms. Extremely high sensitivity in fluorescence detection of many atoms has been shown by spatially resolving each atom in an optical lattice~\cite{Nelson:2007ks,Bakr:2009bx,Sherson:2010hg}. While these systems, with high photon-collection efficiency and long lifetimes can image and count individual atoms in large ensembles, they do not determine the exact atom number. Due to light-assisted collisions in the strongly-confining lattice sites, all atom pairs are lost immediately at the outset of the fluorescence detection. Another promising approach is cavity based detection of mesoscopic samples~\cite{Zhang:2012gw}, which has shown resolution at the single-atom sensitivity level. Here, however, inhomogeneous coupling of the standing-wave light to the atoms has prevented detection of the exact atom number.

Our approach relies on fluorescence detection in a hybrid magneto-optical trap split by a dipole-barrier. We have previously shown single-atom resolution in a conventional MOT for as many as 1200 atoms~\cite{Hume:2013cy}. In such a system, the deep trapping potential, enabling high fluorescence rates and long lifetimes, can allow for measurements with very high signal-to-noise ratio. Here, we extend this idea to the problem of simultaneous detection of two ensembles. These ensembles could be derived, for example, by Stern-Gerlach separation of a two component quantum gas, or directly as the output of an atom-interferometer in spatial degrees of freedom. We show that the lifetime and scattering rates in each of the two zones is sufficient for single-atom resolution in ensembles of hundreds of atoms. In what follows, we first describe our implementation of the hybrid atom trap. We then review the sources of noise that limit the precision of fluorescence measurements and present a detailed analysis of the noise observed in our system.

\section*{The split magneto-optical trap}

We simultaneously detect the individual atom numbers of two atomic ensembles in a novel hybrid trap, as shown in Fig.~\ref{fig:Figure1}(a). A blue-detuned focused light-sheet is superimposed on a $^{87}$Rb MOT to create a potential barrier between the two sites of the resulting double-well system. Once the atoms are loaded into the split MOT, we perform fluorescence imaging on the D$_2$ line to extract the atom number. In Fig.~\ref{fig:Figure1}(b) discrete peaks in the two fluorescence histograms demonstrate single-atom resolution for the two wells. By properly adjusting the height of the barrier between the two wells, we can clearly observe hopping events, Fig.~\ref{fig:Figure1}(c), where a single atom has gained enough energy to surmount the barrier resulting in an anti-correlated step in the fluorescence between the two sites.

In our trap, the MOT laser beams are red-detuned by a frequency $\delta$, nearly equal to the transition linewidth. The precise determination of the atom numbers in each site requires a large potential barrier that suppresses particle exchange. For a given laser power a higher potential barrier height can be achieved by tuning the frequency of the light-sheet closer to resonance. We chose the D$_1$ transition between $5^2\text{S}_{1/2}$ and $5^2\text{P}_{1/2}$ at $\SI{794}{\nano\meter}$ for the light-sheet and employ a narrowband optical filter at $\SI{780}{\nano\meter}$ in order to avoid stray light on the detected images. The natural linewidth of this transition is $\Gamma = 2 \pi \times \SI{5.7}{\mega\hertz}$ and the saturation intensity, assuming $\pi$-polarisation, is $I_{\text{sat}} = \SI{44.86}{\watt/\square\metre}$. The waists of the light-sheet's elliptical cross section are $w_1 = \SI{6}{\micro\metre}$ and $w_2 = \SI{400}{\micro\meter}$ and the laser power is near $\SI{200}{\milli\watt}$. The corresponding Rabi frequency is $\Omega = \Gamma \sqrt{ I / 2 I_{\text{sat}}}$, where $I$ is the light-sheet intensity at the peak of the potential. If we assume $|\Delta| \gg \Omega$, where $\Delta$ is the detuning of the light-sheet, the barrier height can be expressed as $U_{\text{d}} = \hbar \Omega^2 / 4 \Delta$. With a detuning of $\Delta = 2 \pi \times \SI{13}{\giga\hertz}$, by taking into account the transverse profile of the MOT, we expect an effective barrier height near $\SI{13}{\milli\kelvin}$, much larger than the MOT temperature of $\sim \SI{80}{\micro\kelvin}$. Scattering from the light-sheet can happen at a maximum rate of $\Gamma_{\text{sc}} = \Gamma U_{\text{d}} / \hbar \Delta \approx 2 \pi \times \SI{0.2}{\mega\hertz}$ for an atom at the center of the dipole barrier, although the mean scattering rate is much lower, since the atoms are repelled from the position of highest intensity. In any case, the scattering rate $\Gamma_{\text{sc}}$ is considerably smaller than the scattering rate $R_{\text{sc}}$ from the MOT light, which is close to saturation.

% Beam shaping
The light-sheet, an elliptical Gaussian beam with a large aspect ratio, is generated using an optical setup that allows for the easy optimization of the aspect ratio over a large range. A circular Gaussian beam with a waist of $w_0 = \SI{6}{\milli\meter}$ is focused using an achromatic doublet with a focal length of $f_1 = \SI{100}{\milli\meter}$. Without any beam shaping, we measure a resulting radially symmetric waist of $w_1 = \SI{6.1 \pm 0.4}{\micro\meter}$. In order to produce an elliptically shaped beam in the focal plane we use two cylindrical lenses of focal lengths $f_2 = \SI{200}{\milli\meter}$ and $-f_2$, separated by a distance $d$ and placed before the final focussing lens. This results in an axial offset between the positions of horizontal and vertical foci, giving a dipole barrier with adjustable aspect ratio at the position of the atoms. The larger waist, given by $w_2 \simeq d w_0 f_1/f_2^2$, can be adjusted by varying $d$. Limited laser power favors a small waist in order to increase the intensity, while a larger waist is necessary to realize a high homogeneous barrier over the whole MOT size to prevent unwanted hopping. With this in mind, we chose a horizontal waist of $w_2 = \SI{400}{\micro\meter}$ as the final configuration.

% Pulsing
The Stark shift induced by the light-sheet increases the energy of the ground state, thereby disturbing Doppler cooling, and we observe a reduced lifetime when both beams are on at the same time. To avoid this, the light-sheet beam and the MOT beams are pulsed alternately using acousto-optic modulators at a frequency of $\SI{125}{\kilo\hertz}$, which is faster than the characteristic motional frequencies but slow compared to the lifetime of the excited state. Pulsing the lasers in this way produces an effective trapping potential equivalent to the time-average over one pulse cycle. While this reduces the maximum barrier height and the observed fluorescence rate by the respective duty-cycles of the dipole barrier and the MOT, it is essential for reaching trap lifetimes near those of the MOT alone.

% Imaging system
The imaging system used in the experiment is designed to resolve the two ensembles over a depth of field equal to the size of the MOT. Our objective has a numerical aperture of 0.23, and we use a magnification of $\num{5.17 \pm 0.07}$. We estimate the depth of field to be $\SI{60}{\micro\meter}$. The resolution of the imaging system was measured by fitting an Airy function to the intensity profile of a test image and equals $\SI{4.3}{\micro\meter}$ in object space.

\section*{Noise limits in fluorescence imaging}

In measurements of the atom number $N$, the two limiting contributions to the measurement uncertainty are fluorescence noise and atom loss. While a long integration time $t$ reduces the fluorescence noise, it increases the probability of atom loss during the detection. The optimal time minimizes the signal variance $\sigma^2 = N / \eta R_{\text{sc}} t + N t / 2 \tau$, where $\eta$ is the detection efficiency, $R_{\text{sc}}$ the photon scattering rate and $\tau$ the trap lifetime. Other possible noise sources include fluorescence fluctuations due to noise in the excitation laser and loss due to light-assisted collisions. Assuming Gaussian white noise, the first contributes to the variance as $(\alpha N)^2/t$, where $\alpha$ is a measure for the stability of the atom fluorescence. The second contribution takes the form $\beta N^2 t$, where $\beta$ is the rate of light-assisted collisions. Assuming uncorrelated noise sources, the total signal variance is then given by~\cite{Hume:2013cy}
\begin{align}\label{eq:TotalAtomNoise}
\sigma^2 ={}& \frac{N}{\eta R_{\text{sc}}} t^{-1} + (\alpha N)^2 t^{-1} + \frac{N}{2 \tau} t + \beta N^2 t,
\end{align}
where we have assumed that $t \ll \tau$ and $t \ll \beta N$, conditions that are both valid in all our measurements. To achieve the best measurement fidelity, defined as the probability of determining the exact atom number $N$ present at the beginning of the detection, one has to account for the mean atom loss in the experiment during the detection time $t$. Given a detected number $N'$ the atom number $N$ is inferred by $N = N' (1 + t / 2 \tau + \beta N' t / 2)$.

\section*{Noise analysis for the split MOT}

\begin{figure}[t]
\includegraphics[width=0.5\textwidth]{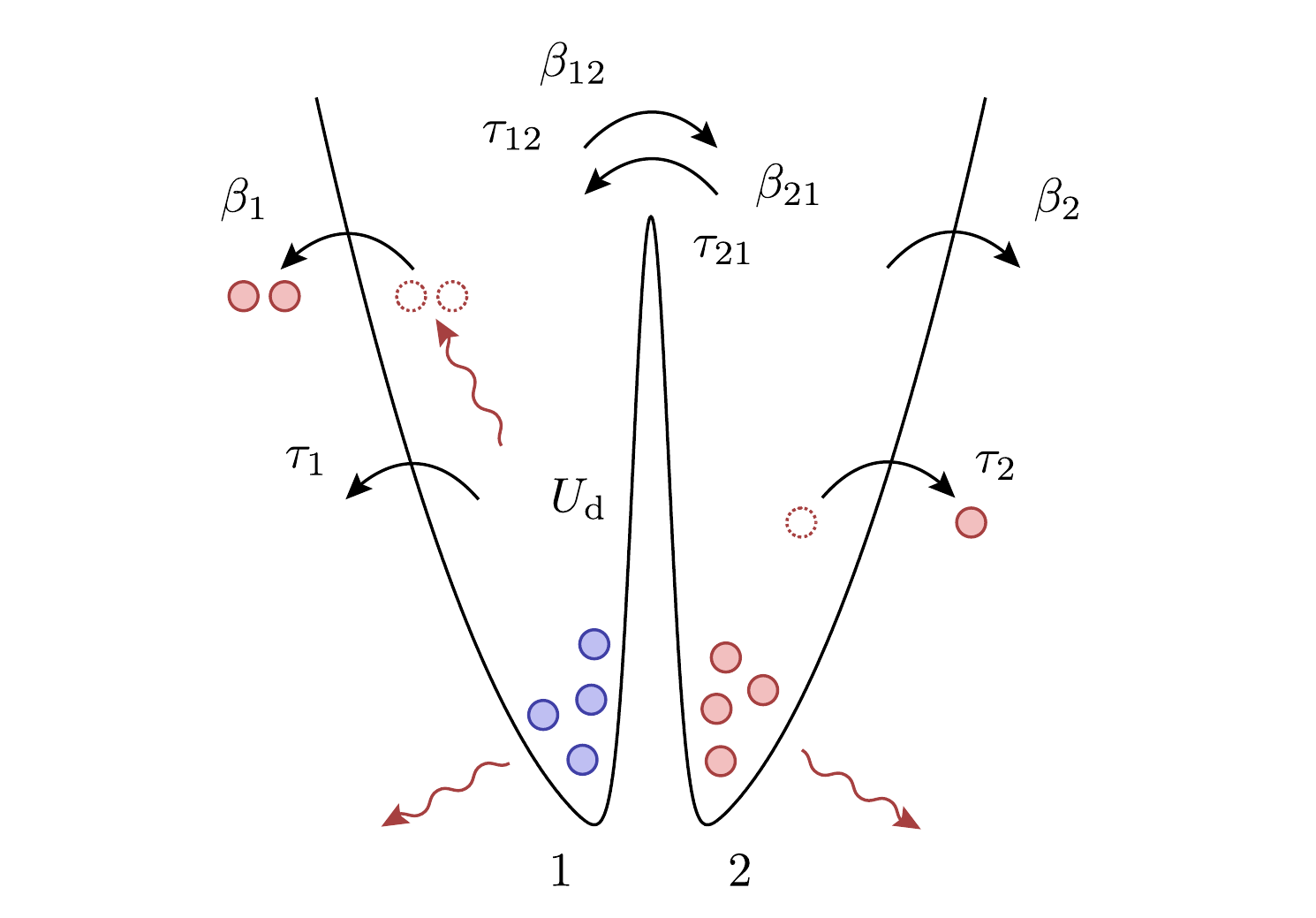}
\caption{(Color online) The atoms reside at the two minima of the dissipative double-well potential. In addition to thermal loss ($\tau_1$, $\tau_2$) and collisional loss ($\beta_1$, $\beta_2$), particle exchange across the potential barrier of height $U_{\text{d}}$ can be initiated by thermal activation ($\tau_{12}$, $\tau_{21}$) and collisional activation ($\beta_{12}$, $\beta_{21}$).}
\label{fig:Figure2}
\end{figure}

% Data acquisition and calibration
For the simultaneous detection of two ensembles the split MOT is imaged onto a low-noise CCD camera and the signal is integrated for typically $\SI{100}{\milli\second}$. Two regions of interest are used to determine the individual fluorescence signals. An example of observed histograms for small numbers of atoms, $N_1$ and $N_2$, are shown in Fig.~\ref{fig:Figure1}(b). For this data, the number of counts per atom, which depends on the overall detection efficiency and may vary with the alignment, is measured to be $\SI{62900\pm200}{\per\second}$ for site 1 and $\SI{63900\pm100}{\per\second}$ for site 2. We have confirmed the linearity of the camera signal at the level of 0.02\% up to hundreds of atoms. Example time traces for an integration time of $\SI{400}{\milli\second}$ are shown in Fig.~\ref{fig:Figure1}(c).

% Denomination of processes
In order to quantify the detection noise for mesoscopic particle numbers we analyse the two-sample variance $\text{Var}(S_{n+1} - S_n)/2$, where $S_n$ and $S_{n+1}$ are consecutively integrated signals. The noise model for the total atom number, given in Eq.~\ref{eq:TotalAtomNoise}, needs to be extended to the simultaneous measurement of two individual atom numbers, taking into account particle exchange between the sites. We now consider the number of particles $N_1$ and $N_2$ in site 1 and 2, respectively, and quantify the rates of loss due to collisions with the background gas by the lifetimes $\tau_1$ and $\tau_2$. Loss due to light-assisted collisions is described by the rates $\beta_1$ and $\beta_2$. Additionally, atoms can hop from one site to the other, either by thermal activation, with mean duration between hopping events denoted as $\tau_{12}$ and $\tau_{21}$, or in the process of a light-assisted collision ($\beta_{12}$ and $\beta_{21}$). All the processes are illustrated in Fig.~\ref{fig:Figure2}.

\begin{figure}[b]
\includegraphics[width=0.5\textwidth]{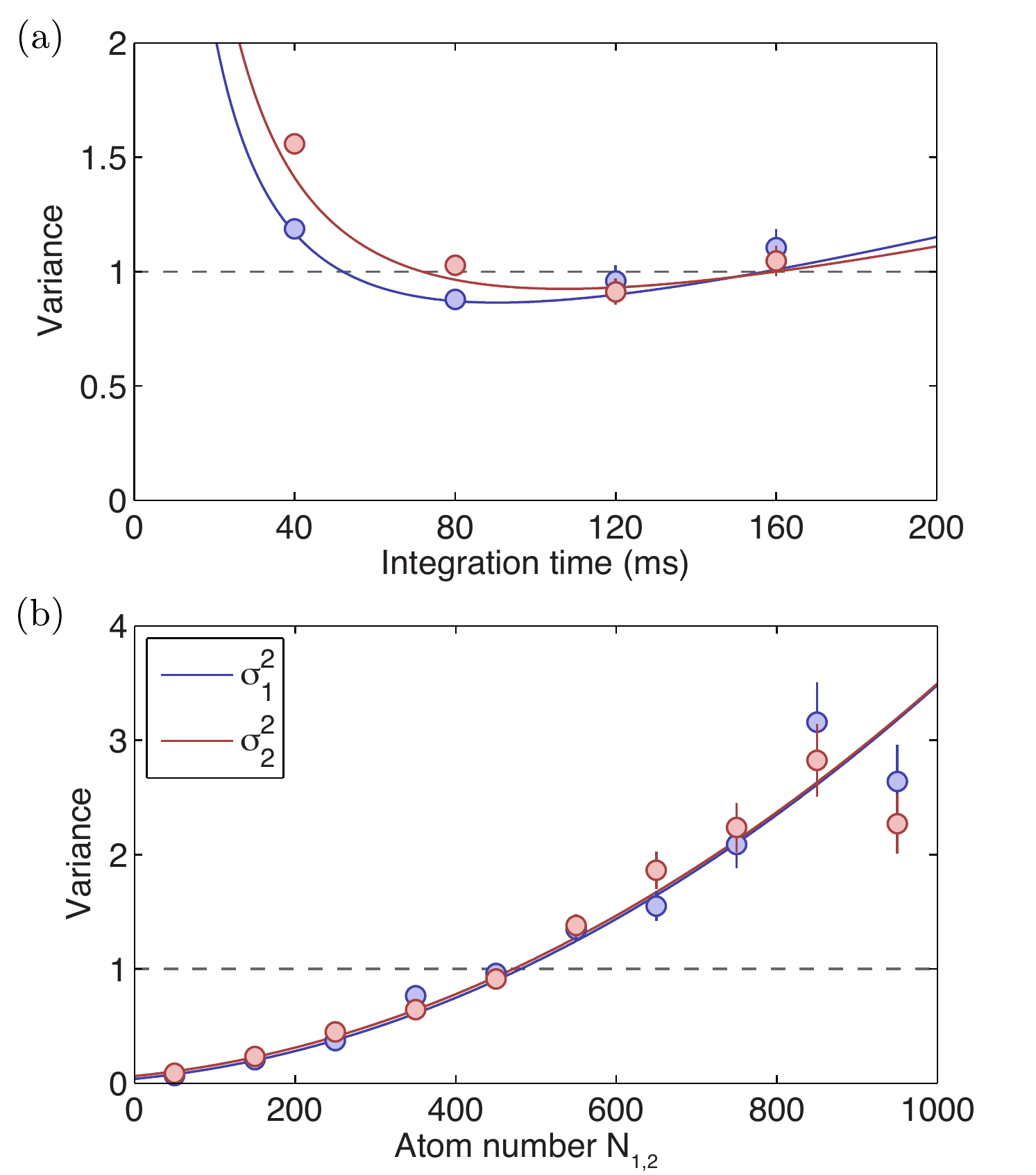}
\caption{(Color online) Variance analysis of individual sites. (a) The variance, shown here for about 450 atoms in each site, reveals a minimum for an optimal integration time between $\SI{80}{\milli\second}$ and $\SI{120}{\milli\second}$. The solid lines are representations of simultaneous fits to a range of integration times and atom numbers. (b) For an integration time of $\SI{120}{\milli\second}$, the variance in each site reaches the single-particle resolution limit $\sigma^2_i = 1$ at 470 atoms.}
\label{fig:Figure3}
\end{figure}

% Noise model for individual sites
The change in the atom numbers $N_1$ and $N_2$ can be described by
\begin{align}
\begin{pmatrix}
\dot{N}_1\\
\dot{N}_2
\end{pmatrix}
={}&
\begin{pmatrix}
-(\tau_1^{-1} + \tau_{12}^{-1}) & \tau_{21}^{-1}\\
\tau_{12}^{-1} & -(\tau_2^{-1} + \tau_{21}^{-1})
\end{pmatrix}
\begin{pmatrix}
N_1\\
N_2
\end{pmatrix}\notag\\
& +
\begin{pmatrix}
-(\beta_1 + \beta_{12}) & \beta_{21}\\
\beta_{12} & -(\beta_2 + \beta_{21})
\end{pmatrix}
\begin{pmatrix}
N_1^2\\
N_2^2
\end{pmatrix},
\end{align}
where the first and second terms account for one- and two-body dynamics, respectively. In analogy to Eq.~\ref{eq:TotalAtomNoise} the variances in site 1 and 2 can be expressed as
\begin{align}
\sigma^2_1 ={}& \frac{N_1}{\eta R_{\text{sc}}} t^{-1} + (\alpha N_1)^2 t^{-1} + \frac{N_1}{2 \tau_1} t + \frac{N_1}{2 \tau_{12}} t + \frac{N_2}{2 \tau_{21}} t\notag\\
& + \beta_1 N_1^2 t + \beta_{12} N_1^2 t + \beta_{21} N_2^2 t
\end{align}
and
\begin{align}
\sigma^2_2 ={}& \frac{N_2}{\eta R_{\text{sc}}} t^{-1} + (\alpha N_2)^2 t^{-1} + \frac{N_2}{2 \tau_2} t + \frac{N_2}{2 \tau_{21}} t + \frac{N_1}{2 \tau_{12}} t\notag\\
& + \beta_2 N_2^2 t + \beta_{21} N_2^2 t + \beta_{12} N_1^2 t,
\end{align}
where we have assumed that the photon shot noise parameter $\eta R_{\text{sc}}$ and the fluorescence noise parameter $\alpha$ are independent of the site. We see that the variance in one site depends on the atom number in the adjacent one. In order to simplify the model and extract the relevant experimental parameters, we consider the case in which $N_1 \approx N_2$. Furthermore, we introduce a noise term $\gamma t^{-1}$ into our model. This term, independent of the atom number, quantifies noise due to stray light, which averages down with increasing integration time. For $i = 1,2$ we obtain
\begin{equation}
\sigma^2_i = \frac{N_i}{\eta R_{\text{sc}}} t^{-1} + \gamma t^{-1} + (\alpha N_i)^2 t^{-1} + \frac{N_i}{2 \tilde{\tau}_i} t + \tilde{\beta}_i N_i^2 t,
\end{equation}
with $\tilde{\tau}_i^{-1} = \tau_i^{-1} + \tau_{12}^{-1} + \tau_{21}^{-1}$ and $\tilde{\beta}_i = \beta_i + \beta_{12} + \beta_{21}$. Based on an independent calibration, for this data, we fix $\eta R_{\text{sc}} = \SI{58496}{\second^{-1}}$ and fit the noise model to the experimentally obtained two-sample variance $\text{Var}(S_{n+1} - S_n)/2$. The fit is performed simultaneously for a range of four different integration times and ten different mean atom numbers. Fig.~\ref{fig:Figure3}(a) shows a representation of this fit for a fixed atom number of 450. We find an optimal integration time between $\SI{80}{\milli\second}$ and $\SI{120}{\milli\second}$. In Fig.~\ref{fig:Figure3}(b) we plot the variance as a function of the atom number for a fixed integration time of $\SI{120}{\milli\second}$ in order to find the single-particle resolution limit $\sigma^2_i = 1$. Since the atom numbers $N_1$ and $N_2$ were kept nearly equal in the experiment, the noise is similar in both sites and we find a limit for single-particle resolved detection of up to 470 atoms in each site.

\begin{table*}[t]
\caption{Fit parameters for the different noise models.}
\begin{ruledtabular}
\begin{tabular}{lllll}
Noise model & $\gamma$ [s] & $\alpha$ [s$^{1/2}$] & $\tilde{\tau}$ [s] & $\tilde{\beta}$ [s$^{-1}$]\\
\\ [-1.5ex] \hline \\ [-1.5ex]
$\sigma^2_1$ & \num{4.6\pm0.7e-3} & \num{3.6\pm0.3e-4} & \num{120\pm52} & \num{1.4\pm0.7e-5}\\
$\sigma^2_2$ & \num{7.6\pm1.2e-3} & \num{4.1\pm0.4e-4} & \num{110\pm67} & \num{1.1\pm1.0e-5}\\
\\ [-1.5ex] \hline \\ [-1.5ex]
$\sigma^2_+$ & \num{1.2\pm0.2e-2} & \num{2.6\pm0.2e-4} & \num{120\pm59} & \num{5\pm30e-7}\\
$\sigma^2_-$ & \num{1.2\pm0.2e-2} & \num{2.9\pm0.3e-4} & \num{100\pm51} & \num{1.2\pm0.5e-5}\\
\end{tabular}
\end{ruledtabular}
\label{tab:NoiseParameters}
\end{table*}

\begin{figure}[b]
\includegraphics[width=0.5\textwidth]{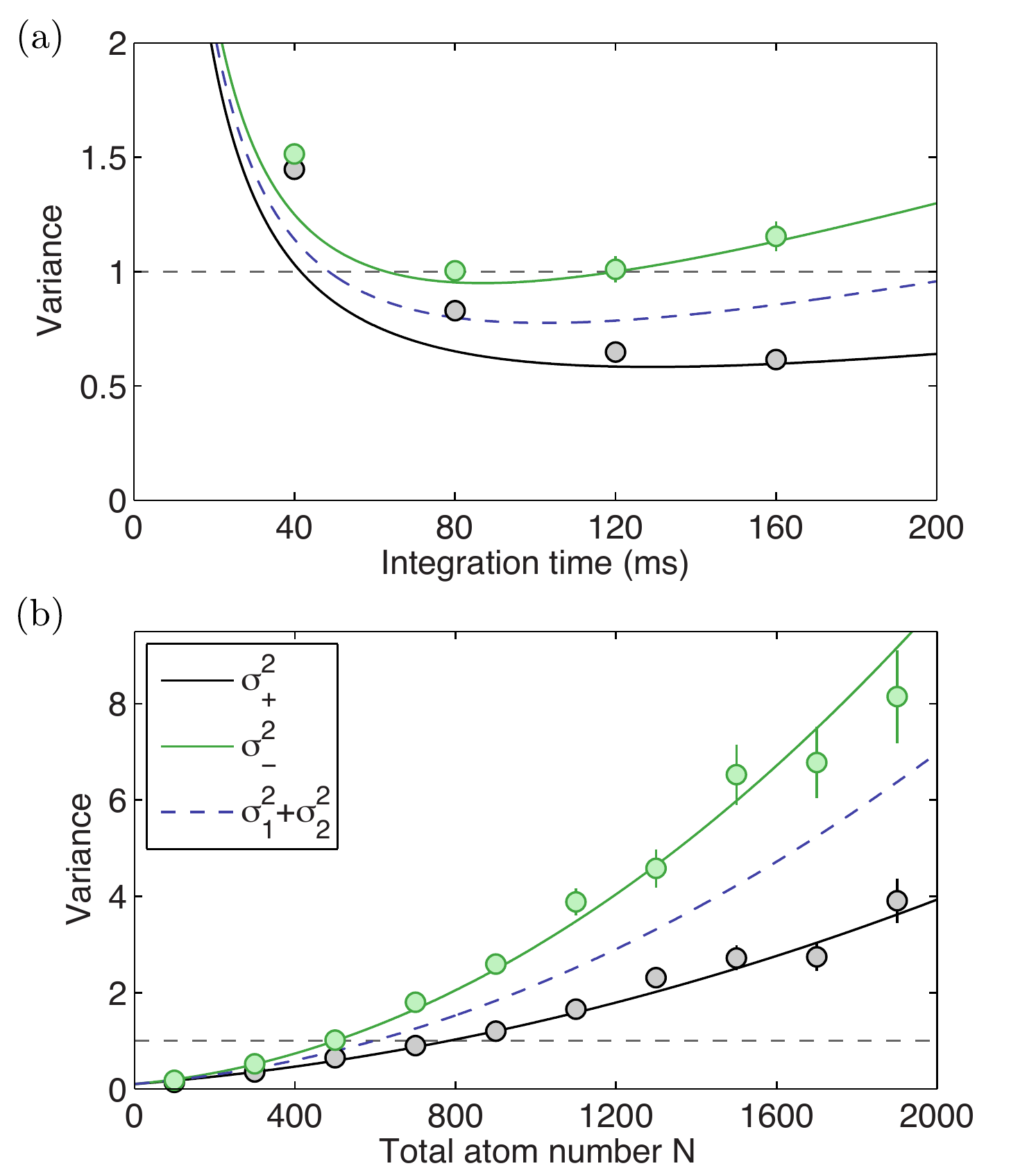}
\caption{(Color online) Analysis of sum and difference variances. (a) We find different optimal integration times for measuring the total atom number $N = N_1 + N_2$ and the atom number difference $N_1 - N_2$, shown here for a mean total atom number of 500. (b) For an integration time of $\SI{120}{\milli\second}$, optimised for measuring the total atom number, we find the single-particle resolution limit at 500 atoms for measuring $N_1 - N_2$ (green upper line) and close to 800 atoms for measuring $N$ (black lower line). In between lies the sum of the individual variances (blue dashed line).}
\label{fig:Figure4}
\end{figure}

% Noise model for atom number sum and difference
For many measurement scenarios we will be interested in the total atom number, $N = N_1 + N_2$, and the atom number difference, $N_1 - N_2$. For the total atom number the noise model reads
\begin{align}
\sigma^2_+ ={}& \frac{N}{\eta R_{\text{sc}}} t^{-1} + \gamma_+ t^{-1} + (\alpha_+ N)^2 t^{-1} + \frac{N}{2 \tau} t + \beta N^2 t.
\end{align}
If we assume decay parameters $\tau \equiv \tau_1 \approx \tau_2$ and $\beta \equiv \beta_1 \approx \beta_2$, as well as exchange parameters $\tau_{\text{ex}} \equiv \tau_{12} \approx \tau_{21}$ and $\beta_{\text{ex}} \equiv \beta_{12} \approx \beta_{21}$, we can express the total atom number variance as $\sigma^2_+ = \sigma^2_1 + \sigma^2_2 + 2 \text{Cov}(N)$ with the atom covariance
\begin{equation}\label{eq:AtomCovariance}
\text{Cov}(N) \equiv -\frac{1}{2} \left( \frac{N}{\tau_{\text{ex}}} + \beta_{\text{ex}} N^2 \right) t.
\end{equation}
The atom covariance is always negative, since particle exchange events have an anti-correlated effect on $N_1$ and $N_2$. This behavior can be observed in Fig.~\ref{fig:Figure1}(c). The variance of an atom number difference measurement can be expressed as $\sigma^2_- = \sigma^2_1 + \sigma^2_2 - 2 \text{Cov}(N)$. Both sum and difference variances, as well as the sum of the individual variances, are shown in Fig.~\ref{fig:Figure4}. Since $\sigma^2_-$ is always larger than $\sigma^2_+$ due to noise from particle exchange, the measurement of the atom number difference favours a shorter integration time of $\SI{80}{\milli\second}$ compared to the optimal integration time of $\SI{120}{\milli\second}$ for the sum measurement. We find a single-particle resolution limit of $\sigma^2_+ = 1$ for a total atom number of 800 and $\sigma^2_- = 1$ for 500 particles. The detection performance can be expressed as a measurement fidelity, defined as the probability of detecting exactly the initial atom number. For $N = 100$, the fidelity of the sum measurement is 87.4\%, while for the difference measurement it is 87.3\%. The reduced performance compared to previous measurements in a single MOT~\cite{Hume:2013cy} is mainly due to increased fluorescence noise.

% Fit parameter discussion
Table~\ref{tab:NoiseParameters} shows the parameters obtained from the different noise models. Comparing the results for $\sigma^2_1$ and $\sigma^2_2$ we find that, within the error of the measurement, $\alpha$ is indeed independent of the site, as expected. The same holds for the one-body parameter $\tilde{\tau}$ and the two-body parameter $\tilde{\beta}$. For the derivation of Eq.~\ref{eq:AtomCovariance} we have used $\gamma_+ = 2 \gamma$ and $\sqrt{2} \alpha_+ = \alpha$. This is confirmed by the obtained fit parameters. For $\sigma^2_+$ the one-body parameter $\tilde{\tau}$ and coincides with the appropriate parameter in $\sigma^2_i$, corresponding to the lifetime of the trap. From this we deduce that our maximum potential barrier height is large enough, such that thermal hopping between the two sites of the double-well is negligible, and the one-body limitation is solely given by the loss from the trap. The situation is different for dynamics due to light-assisted collisions. From $\sigma^2_+$ we obtain a two-body parameter which is orders of magnitude smaller than the $\tilde{\beta}$ parameter in $\sigma^2_i$ and $\sigma^2_-$, where hopping activated by collisions dominates the loss. With the given barrier height we are in a regime in which thermal hopping between the sites is strongly suppressed, however, light-assisted collisions, due to their longer-ranged exponential energy distribution still contribute significantly.

\section*{Conclusion}

In summary, we have shown simultaneous determination of the total atom number with single-atom resolution in two spatially separated mesoscopic samples -- a prerequisite for achieving Heisenberg-limited interferometry. The hybrid trap, consisting of a dipole barrier superimposed on a MOT is designed for a high fluorescence rate and long trap lifetime, enabling fluorescence measurements with high signal-to-noise ratios. We use a model that includes all known sources of noise. Fits of this model to experimental noise measurements yield a set of parameters describing the particle loss and exchange rates, both due to collisions with background gas and light-assisted collisions, as well as fluorescence noise parameters.  By comparing fits for the two individual zones, the atom number sum and difference we find these parameters are internally consistent and match separate calibrations where available. Independent of the accuracy of the noise model, we have directly measured a single-particle resolution limit for detecting the atom number difference at a total of 500 atoms in the ensemble.  

Since we have chosen a detection time that minimizes the measured variance, the remaining noise is equally due to fluorescence noise and atom loss. Both of these can in principle be further reduced. In the case of fluorescence noise we have not reached the photon shot noise limit, so more careful stabilization of the laser frequency and intensity may yield an improvement. We find that the loss and exchange rates include important contributions from both a linear loss process, presumably due to collisions with background gas, and light-assisted collisions. To significantly reduce the linear loss it would likely be necessary to reduce the vacuum pressure in our chamber. The light assisted collisions might be further reduced by lowering the trap density.

Our general strategy of dividing a MOT into separate zones using a dipole barrier can be applied in a straightforward way to different atomic species and more complicated multi-zone trap geometries. For state-selective detection of a two-component Bose gas, we plan to first separate the two magnetic sublevels with a Stern-Gerlach pulse, and load them into the two trap zones for fluorescence detection. If this can be performed with high fidelity, our system could be used to significantly reduce detection noise in existing spin-squeezing experiments~\cite{Strobel:2014eg}. An exact atom counter for the two spin states will allow for the realization of an atomic analog to the $N$-particle Hong-Ou-Mandel experiment~\cite{LewisSwan:1ig} from quantum optics.

\section*{Acknowledgments}

\begin{acknowledgments}
The authors would like to thank Helmut Strobel, Wolfgang Muessel, and Daniel Linnemann for helpful discussions. This work was supported by the Heidelberg Center for Quantum Dynamics, Forschergruppe FOR760 of the Deutsche Forschungsgemeinschaft, and the FET-Open project QIBEC (Contract No. 284584). I.S. acknowledges support from the International Max Planck Research School (IMPRS-QD), and D.B.H. from the Alexander von Humboldt Foundation.
\end{acknowledgments}

\bibliography{Paper}

\end{document}